\documentclass[preprint,showpacs,aps,floatfix]{revtex4} 
\usepackage{graphicx}
 
\begin{document}  

\title{Stochastic Simulations of the Repressilator Circuit} 

\author{Adiel Loinger$^1$ and Ofer Biham$^1$
}  
\affiliation{  
$^1$ Racah Institute of Physics,   
The Hebrew University,   
Jerusalem 91904,   
Israel
}  
 
\begin{abstract} 
 
The genetic repressilator circuit consists of three 
transcription factors, or repressors, which negatively 
regulate each other in a cyclic manner. This circuit was 
synthetically constructed on plasmids in {\it Escherichia coli} 
and was found to exhibit oscillations in the concentrations 
of the three repressors. Since the repressors and their 
binding sites often appear in low copy numbers, the 
oscillations are noisy and irregular. Therefore, the 
repressilator circuit cannot be fully analyzed using 
deterministic methods such as rate-equations. Here we 
perform stochastic analysis of the repressilator circuit 
using the master equation and Monte Carlo simulations.
It is found that fluctuations modify the range of conditions
in which oscillations appear as well as their amplitude and 
period, compared to the deterministic equations. The deterministic 
and stochastic approaches coincide only in the limit in which all the relevant 
components, including free proteins, plasmids and bound proteins,
appear in high copy numbers. We also find that subtle 
features such as cooperative binding and bound-repressor degradation
strongly affect the existence and properties of the oscillations.

\end{abstract}

 
\maketitle  

\section{Introduction}
\label{sec:Introduction}

Regulation processes in cells are performed by networks
of interacting genes, which regulate each other's synthesis.
In recent years these networks have been studied extensively
in different organisms
\cite{Alon2006,Palsson2006}.
The networks include interactions at the level of transcriptional
regulation 
\cite{Milo2002,Milo2004}
as well as post-transcriptional regulation by
protein-protein interactions
\cite{Yeger-Lotem2004}.
In attempt to understand the structure of the networks and their
function, it was proposed that they exhibit a 
modular structure~\cite{Milo2002,Milo2004,Yeger-Lotem2004}
with motifs, such as the 
feed forward loop~\cite{Mangan2003}.
Other modules such as the
genetic switch~\cite{Ptashne1992} and 
the mixed-feedback loop~\cite{Yeger-Lotem2004,Francois2005}
also appear.
However, it is not yet clear what is the connection between
the evolutionary processes which determine the network structure 
and the functionality of these motifs
\cite{Artzy2004,Mazurie2005,Meshi2007}.

In addition to the genetic circuits found in natural
organisms, it recently became possible to construct synthetic
networks of a desired architecture
\cite{Gardner2000,Elowitz2000}.
An important example of a synthetic circuit is the
repressilator~\cite{Elowitz2000}, which was designed to exhibit
oscillations, reminiscent of natural genetic
oscillators such as the circadian rhythms.
The repressilator circuit was encoded on plasmids
in {\it E. coli} bacteria.
The plasmids
appeared in a low copy number of about
four plasmids per cell. 
The reporter plasmid transcribing the GFP used for
the measurements appeared in a higher copy number of 
around 65 plasmids per cell. 
The protein concentrations
were measured vs. time in single cells.
Oscillations with a period of  
160 $\pm$ 40
minutes 
were found. 
Note that this oscillation period 
was longer than the cell cycle, 
of about 50-70 minutes under the experimental conditions.
The oscillations were 
noisy, typically maintaining phase coherence for 
only a few oscillation periods. 
In addition, the reporter gene expression 
exhibited a rising background level
as time evolved. 

\begin{figure}
\centering
\includegraphics[width=6cm]{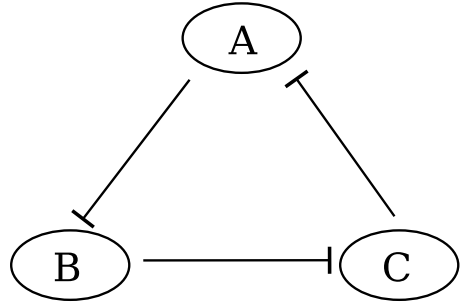}
\caption{
Schematic plot of the repressilator circuit.
It consists of three genes which 
negatively regulate each other 
by transcriptional regulation
in a cyclic order.}
\label{fig:1}
\end{figure}

The repressilator circuit 
consists of three genes, denoted by $a$, $b$ and $c$, 
which negatively regulate each other's synthesis
in a cyclic fashion,
namely $a$ regulates $b$, $b$ regulates $c$ and $c$ 
regulates $a$
(Fig.~\ref{fig:1}).
The regulation is performed by the transcription factors,
or repressors, 
$A$, $B$ and $C$, produced by genes $a$, $b$ and $c$,
respectively. When a repressor binds
to the promoter site upstream of the regulated gene, 
it blocks the access of the RNA polymerase, 
thus repressing the transcription process.

To understand the oscillatory behavior, 
consider a situation in which the number of $A$ proteins is large. 
In this case it is likely that one of the $A$ proteins
will bind to the $b$ promoter and will 
repress the production of $B$ proteins.
The reduced level of $B$ proteins will enable the gene
$c$ to be fully expressed and the number of $C$ proteins
will increase and will start to repress gene $a$.
As a result, the number of $A$ proteins will decrease,
and gene $b$ will be activated, completing a full cycle,
in which the order of appearance of the dominant protein
type is $A \to C \to B \to A$. 

In this paper we analyze the repressilator circuit
using deterministic methods (rate-equations)
and stochastic methods 
(direct numerical integration of the master equation and 
Monte Carlo simulations).
Recent advances in quantitative measurements of protein levels
in single cells~\cite{Elowitz2002,Ozbudak2002}
gave rise to new insight into the importance
of stochastic fluctuations~\cite{Mcadams1997,Mcadams1999,Paulsson2004}.
The role of fluctuations is enhanced due to the discrete nature
of the transcription factors and their binding sites,
which may appear in low copy numbers~\cite{Becskei2000,Kaern2005}.
Using stochastic methods we examine the effect of
fluctuations on the regularity, amplitude and frequency
of the oscillations.
In particular, we examine the effect of the number 
of binding sites by changing the number of plasmids in a cell.
We find that when the number of plasmids in small,
fluctuations are important and stochastic analysis is
required. In the limit of a large number of plasmids
the fluctuations decline and the deterministic
and stochastic results coincide.
We also consider the effects of features such as
cooperative binding,
the inclusion of the mRNA level in
the models and bound-repressor degradation.
The appearance of oscillations turns out to be sensitive to 
such features and it is thus essential to study them
in detail. The results provide 
concrete predictions
for systematic experimental studies
using plasmids.

The paper is organized as follows.
In Sec. II we review the commonly used 
model of the repressilator circuit, 
based on the Michaelis-Menten kinetics.
In Sec. III we present a more complete deterministic analysis of
the repressilator using rate-equations. 
In Sec. IV we present a stochastic analysis 
and examine the effect of fluctuations
on the appearance and regularity of the oscillations
as well as on their amplitude and frequency.
The differences between the deterministic and stochastic 
results and the effects of the number of binding sites
are discussed in Sec. V. 
The results
are summarized in Sec. VI.

\section{Michaelis-Menten Rate-Equation Model}

Following Ref.~\cite{Elowitz2000},
we first analyze 
the repressilator circuit using the standard Michaelis-Menten
rate-equations. 
These equations describe the time evolution of
the concentrations of the various proteins and mRNA molecules in the cell.
By concentration of a certain molecule we refer here to 
its average copy number per cell. 
For simplicity we denote the three proteins by
$X_1 = A$, $X_2 = B$ and $X_3 = C$, and the corresponding
mRNAs by $m_i$.
The concentration of the free $X_i$ protein $(i=1,2,3)$ is given
by $[X_i]$, while the concentration of the corresponding mRNA
is given by $[m_i]$.

In the Michaelis-Menten equations, the fact that 
the transcription of protein $B$ is 
negatively regulated by protein $A$, is taken into account by reducing
the transcription rate of protein $B$ by a factor of $1+k[A]^n$, 
which is called the Hill-function.
In this expression, $k$ is a parameter that quantifies the repression
strength (or the affinity between the transcription factor and the promoter).
The parameter $n$ is called the Hill-coefficient.
Hill-function models are simplifications of rate-law equations.
When derived directly from rate laws, 
$n$ is expected to take non-negative 
integer values.
In this case, $n$ 
represents the number of transcription
factors required 
to bind simultaneously in order
to perform the regulation. 
However, when these models are used for fitting 
empirical data, $n$ is considered as
a fitting parameter which may take 
non-integer values. 
In the analysis below, we consider only 
non-negative
integer values of $n$.
The Michaelis-Menten equations for the repressilator are
%
%
%
\begin{equation}
\left\{
\begin{array}{l}    
\dot{[m_i]} = \frac{g_m}{1+k[X_{i-1}]^n} - d_m [m_i]  \\
\dot{[X_i]} = g_p [m_i] - d_p [X_i],
\end{array}
\right.
\label{eq:michaelis_menten_with_mrna}
\end{equation}    
where $i=1,2,3$. 
Note that the indices form a cyclic set 
$X_i$, $i=1,2,3$, namely $X_0$ is identified as $X_3$.
The transcription and translation rates are $g_m$ and $g_p$ (s$^{-1}$),
respectively. 
The degradation rates of mRNAs and proteins are given by $d_m$ and $d_p$
(s$^{-1}$), respectively.
For simplicity we assume identical parameters for the three proteins.

Often, the mRNA level is ignored and 
the protein is regarded as produced in a single step
of synthesis 
\cite{Becskei2000,Rosenfeld2002,Kepler2001,Sasai2003}.
In this case the effective rate of protein production is
$g=g_p g_m/d_m$ and the Michaelis-Menten equations are
\begin{equation}    
\dot{[X_i]} = \frac{g}{1+k[X_{i-1}]^n} - d [X_i],
\label{eq:michaelis_menten_without_mrna}
\end{equation}    
where $d=d_p$.
Ignoring the mRNA level may be justified under steady state conditions. 
However, when oscillations take place, 
the mRNA level may be important.
Including the mRNA may account for an effective delay
in the production of the protein.
This observation is supported by the fact that
delays can be approximated by adding certain intermediate
steps to the dynamical model
~\cite{Mocek2005}.
Such delays have been shown to have importance in the emergence of
oscillations 
\cite{Chen2002,Lewis2003,Monk2003,Bratsun2005}.

The Michaelis-Menten equations presented above exhibit 
a single steady-state solution for any choice of the parameters.
However, in some range of parameters this solution may become unstable 
and oscillations emerge.
It turns out that the conditions for oscillations depend on
the Hill-coefficient.
For Hill-coefficient $n=1$ no oscillations appear.
For $n=2$, the system oscillates (for suitable parameters)
in case that the mRNA level is included, 
but does not oscillate in case that it is ignored.
For $n=3$, the system exhibits oscillations even
if the mRNA level is ignored (Table I).
These results indicate that oscillations are favored by
high nonlinearity or delays, in agreement with Ref.
\cite{Griffith1968}.

\section{Deterministic Analysis}

\subsection{Repressilator Without Cooperative Binding}

Consider the repressilator circuit 
without cooperative binding, namely with 
Hill-coefficient $n=1$. 
In this case
the regulation of each gene
is performed by a single bound protein.
We will show below that although
the Michaelis-Menten equations do not exhibit oscillations,
a slight modification of the circuit architecture will lead to
oscillations. For the case of $n=1$ we ignore the mRNA level 
because adding it does not change the behavior of the circuit.

The Michaelis-Menten equations presented above 
provide a rather crude description of the transcriptional
regulation process. 
In order to model this process in greater detail 
we introduce below a more complete set of rate-equations.
These equations account for the free repressors and for
the bound repressors as two separate populations.
We denote by $r_i$ those $X_i$ proteins which are bound to
the promoter site, where they perform the regulation process.
In the repressilator circuit,
$r_A$ is a bound $A$ protein that regulates the production of $B$, 
$r_B$ is a bound $B$ protein that regulates the production of $C$, and
$r_C$ is a bound $C$ protein that regulates the production of $A$.
The average number of bound proteins in a cell is denoted
by $[r_i]$, $i=1,2,3$.
Here we consider the case where there is a single gene of each
type and the expression of each gene is regulated by a single
binding site.
Each binding site may be either vacant or 
occupied by a single bound repressor.
When the promoter site of the gene $X_i$ is vacant, the
gene is expressed and proteins are produced at rate $g$.
When the promoter site is occupied (by a bound repressor $r_{i-1}$),
the gene is not expressed and no proteins are produced.
The average production rate of protein $X_i$ will thus be
$g [e_{i-1}]$, where $[e_{i-1}]$
is the average number of vacant binding sites.
Since there is a single binding site for each gene,
it is clear that $[e_i] + [r_i] =1$ for
$i=1,2,3$.
Thus, the production rate 
of protein $X_i$
can be expressed by
$g \cdot (1-[r_{i-1}])$.
The rate-equations for the repressilator circuit will
thus take the form

%
\begin{equation}
\left\{
\begin{array}{l}
\dot{[X_i]} = g \cdot  (1-[r_{i-1}])-d [X_i]
-\alpha_0[X_i]\left(1-[r_i]\right)
+\alpha_1[r_i],  \\
\dot{[r_i]} = \alpha_0[X_i]\left(1-[r_i]\right)-\alpha_1[r_i],
\end{array}
\right.
\label{eq:extended_rate}
\end{equation}
where $i=1,2,3$.
The parameter $\alpha_0$ (s$^{-1}$ molecule$^{-1}$) 
is the binding rate of the 
transcription factors to the promoter site, 
while $\alpha_1$ (s$^{-1}$)
is their un-binding rate.
In the limit where the binding and un-binding processes 
are much faster than the other relevant 
processes in the system, namely 
$\alpha_0,\alpha_1 \gg d,g$,
these equations can be reduced to the Michaelis-Menten form. 
In this limit, the relaxation times of $[r_i]$ 
are much shorter than other relaxation times in the system. 
Thus, one can take the 
time derivatives of $[r_i]$ to zero, 
even if the system is away from steady state.
This brings the rate-equations to the 
Michaelis-Menten form [Eq.~(\ref{eq:michaelis_menten_without_mrna})]
with $n=1$ and $k=\alpha_0/\alpha_1$.

Eqs.~(\ref{eq:extended_rate}) exhibit
a single positive steady state solution
\begin{equation}
\label{eq:repressillator_steady_state_1}
[X_i] = \frac {-1+\sqrt{1+4kg/d}}{2k}, \ \ i=1,2,3.
\end{equation}
Linear stability analysis 
shows that
this solution is stable for any choice of the parameters.
Therefore, this circuit cannot sustain oscillations
(although it may exhibit damped oscillations).
Including the mRNA level in the equations does not change this result,
as long as $n=1$.

Unlike the Michaelis-Menten approach, 
Eqs.~(\ref{eq:extended_rate})
include a separate population of bound repressors.
This enables us to consider the possibility that bound
repressors degrade. 
Although the degradation of bound transcription factors 
is not commonly discussed in the biological literature,
it may have biological relevance.
Moreover, some theoretical models 
implicitly assume that bound proteins degrade at the 
same rate as free proteins
\cite{Hornos2005,Kim2007}.
Note that even without degradation of bound repressors at the
molecular level, cell division introduces an effective degradation of all
proteins including bound transcription factors. 
This is due to the fact that
during the DNA replication only one of the 
two DNAs will have a repressor attached to it.
It turns out that bound-repressor degradation (BRD)
gives rise to oscillations even without cooperative binding,
regardless of whether the mRNA level is included or not.
This result is valid even when the degradation rate for bound
repressors is significantly lower than for free repressors.

It should be noted that the degradation of a bound repressor is
fundamentally different from the unbinding of such repressor.
Degradation removes the repressor from the system, while 
unbinding enables the repressor to bind again.
This difference is most crucial when the repressor appears in
a low copy number. If the degradation of bound repressors is
not taken into account, the last repressor may repeatedly
bind and unbind, being bound most of the time. 
As a result, its effective degradation rate is significantly
reduced.

\begin{figure}
\includegraphics[width=14cm]{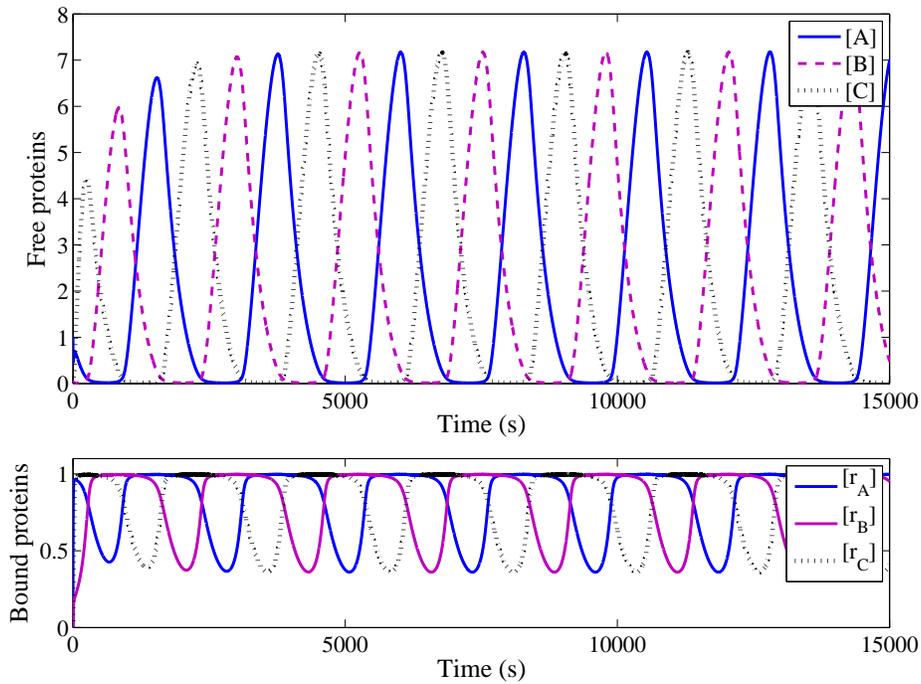}
\caption{
(color online)
The numbers of free proteins (top) 
and bound proteins (bottom) of type $A$, $B$ and $C$, vs. time,
for the repressilator with bound repressor degradation,
obtained from numerical integration of the rate-equations.
The parameters used are:
$g=0.05$, $d=d_r=0.003$, $\alpha_0=0.5$ 
and $\alpha_1=0.01$ (s$^{-1}$).
A regular pattern of oscillations is observed, 
with a phase delay of $120^0$ between successive proteins.
The period of the oscillations is $\simeq 2260$ (s),
and the maximal copy number is $\simeq 7$. 
}
\label{fig:2}
\end{figure}

Denoting the degradation rate of the bound repressors 
by $d_r$ (s$^{-1}$), we obtain the following rate-equations
%
%
%
\begin{equation}
\left\{
\begin{array}{l}
\dot{[X_i]} = g \cdot (1-[r_{i-1}])-d [X_i]-\alpha_0[X_i]\left(1-[r_i]\right)
+\alpha_1[r_i] \\
\dot{[r_i]} = \alpha_0[X_i]\left(1-[r_i]\right)-\alpha_1[r_i]-d_r[r_i].
\end{array}
\right.
\label{eq:rate_brd}
\end{equation}
These equations exhibit oscillations for a broad range of parameters,
and specifically for a broad range of values of $d_r$.
These oscillations, 
shown in Fig.~\ref{fig:2},
are clearly non-sinusoidal.
Indeed, the order of appearance of the dominant protein species
is $A \to C \to B \to A$, as expected. 
The oscillations are symmetrical in the sense that 
the oscillation patterns for the three proteins are
identical and each protein is dominant during about $1/3$ of the cycle.
When different parameters are chosen for the three proteins, 
the amplitudes of their oscillations become different. 
Also, a protein that exhibits a larger amplitude
maintains its dominance for a larger fraction of the
oscillation period. 

The parameter range in which oscillations are present
shrinks to zero when $d_r \rightarrow 0$.
We have analyzed the dependence of the oscillation
period and amplitude on the parameters.
It was found that the oscillation period, $T$, is dominated
by the degradation rate of the proteins, namely $T \sim 1/d$.
Since the lowest value of $[X_i]$ during the oscillation is 
typically nearly zero, the amplitude is given by 
$[X_{\rm max}]/2$,
where
$[X_{\rm max}] \sim g/d$
is the largest value of 
$[X_i]$.

As before, in the limit of fast binding and unbinding, 
one can obtain, 
from Eqs.~(\ref{eq:rate_brd}), 
modified 
Michaelis-Menten equations of the form:
\begin{equation}    
\dot{[X_i]} = \frac{g}{1+k[X_{i-1}]} 
- d [X_i]-{{d_rk}\over{1+k[X_{i}]}} [X_i].
\label{eq:michaelis_menten_brd}
\end{equation}

\noindent    
These equations also exhibit oscillations, 
unlike the ordinary Michaelis-Menten Equations.
The oscillations are very similar to those
obtained from 
Eqs.~(\ref{eq:rate_brd}).
The effect of BRD on the oscillations can be understood from
Eq. (\ref{eq:michaelis_menten_brd}).
This equation shows that BRD introduces a nonlinear
degradation term to $[X_i]$. In this term, the degradation rate
decreases as $[X_i]$ increases. This helps to destabilize the
steady state solution. Small deviations 
from the steady state are enhanced because
a protein that appears in small numbers has a higher degradation
rate than a protein that appears in large numbers.

\subsection{Repressilator with Cooperative Binding}

In transcriptional regulation with cooperative binding,
two or more copies of the transcription factor need to 
bind simultaneously to the promoter in order to 
perform the regulation. 
The number of simultaneously bound transcription factors
needed to perform the regulation is given by $n$.
The effect of cooperative binding was studied extensively 
before in the context of the genetic toggle switch, which 
consists of two genes which negatively regulate each
other's synthesis
\cite{Cherry2000,Warren2004,Warren2005,Walczak2005,Lipshtat2006,Loinger2007}.
It was found to have important effects on the function 
and stability of the genetic switch.

Here we focus on the repressilator circuit in the case of $n=2$.
In particular, we consider the case in which
pairs of identical proteins bind to each other and form dimers,
namely,
$X_i + X_i \rightarrow D_i$.
When such dimer binds to a suitable promoter site, 
it negatively regulates the expression of the corresponding gene.
In the analysis below we also account for
the mRNA level, considering the transcription and translation
as two separate processes.
The rate-equations describing the repressilator system are

%
\begin{equation}
\left\{
\begin{array}{l}
\dot{[m_i]} = g_m \cdot (1 - [r_{i-1}]) - d_m [m_i] \\
\dot{[X_i]} = g_p [m_i] - d [X_i] - 2\gamma [X_i]^2 \\
\dot{[D_i]} = \gamma [X_i]^2 
- d_D [D_i] - \alpha_0 [D_i] (1-[r_i]) +\alpha_1 [r_i] \\
\dot{[r_i]} = \alpha_0 [D_i] (1-[r_i]) - \alpha_1 [r_i], 
\end{array}
\right.
\label{eq:rate_dimers}
\end{equation}
where $\gamma$ is the dimerization rate constant 
and $d_D$ is the degradation rate of the dimers.
These equations exhibit oscillations within some range of parameters. 
We find that within the deterministic framework,
including the mRNA level is sufficient
in order to obtain oscillations. 
However, even if the mRNA level is ignored,
oscillations take place if bound-repressor degradation
is taken into account (Table II).

\section{Stochastic Analysis}

\subsection{Repressilator without Cooperative Binding}

To account for stochastic effects we analyze the 
repressilator system using
the master equation
\cite{Paulsson2000,Kepler2001,Paulsson2002} 
and Monte Carlo (MC) simulations
\cite{Gillespie1977,Mcadams1997,Mcadams1999}.
The role of fluctuations is enhanced due to 
the discrete nature of the transcription factors
and their binding sites, which may appear in low copy numbers.
We also gain insight into the role 
of bound repressor degradation in the emergence of oscillations.

In the stochastic description of the system, 
we denote the number of free $X_i$ proteins by $N_i$,
and the number of bound $X_i$ proteins by $r_i$.
Using the master equation, 
we consider the time evolution of the
probability distribution function $P(N_A,N_B,N_C,r_A,r_B,r_C)$,
or in a more convenient notation $P(N_1,N_2,N_3,r_1,r_2,r_3)$.
This is the probability for a cell to 
include $N_i$ copies of free protein $X_i$
and  $r_i$ copies of the bound $X_i$ repressor,
where $N_i=0,1,2,\dots$,  
and $r_i = 0,1$ 
(assuming a single binding site).
The master equation for the repressilator 
without cooperative binding takes the form
\begin{eqnarray}
\label{eq:oscillator_BRD_master}
&& \dot{P}(N_1,N_2,N_3,r_1,r_2,r_3) = \\
&&   \sum_{i=1,2,3} \lbrace 
 g \cdot (1-r_{i-1}) [P(\dots,N_i-1,\dots,r_1,r_2,r_3) 
- P(N_1,N_2,N_3,r_1,r_2,r_3)] \nonumber\\
&& + d (N_i+1) [P(\dots,N_i+1,\dots,r_1,r_2,r_3) 
- N_i P(N_1,N_2,N_3,r_1,r_2,r_3)] \nonumber\\
&& + \alpha_0 (N_i+1) r_i [P(\dots,N_i+1,\dots,r_i-1,\dots) - 
     N_i (1-r_i)  P(N_1,N_2,N_3,r_1,r_2,r_3)] \nonumber\\
&& + \alpha_1  (1-r_i) [P(\dots,N_i-1,\dots,r_i+1,\dots) - 
     r_i  P(N_1,N_2,N_3,r_1,r_2,r_3)] \nonumber\\
&& + d_r  (r_i+1) [P(N_1,N_2,N_3,\dots,r_i+1,\dots) 
- r_i P(N_1,N_2,N_3,r_1,r_2,r_3)] \rbrace . \nonumber
\end{eqnarray}
The master equation has a single steady state solution,
$\dot P(\vec N)=0$, for all  
$\vec N=(N_1,N_2,N_3,r_1,r_2,r_3)$.
This solution can be obtained by direct numerical integration
of the master equation and it is always stable~\cite{VanKampen1992}.
The steady state solution of this master equation is not
an equilibrium state, and therefore detailed balance
is not satisfied. As a result, there is a net flow of probability
between adjacent $\vec N$ states. The net flux of probability 
between states $\vec N$ and  $\vec N'$ is given by
\begin{equation}
\phi(\vec N \to \vec N') = W(\vec N \to \vec N')P(\vec N) 
- W(\vec N' \to \vec N)P(\vec N'),
\end{equation}
where $W(\vec N \to \vec N')$ is the transition rate from
$\vec N$ to $\vec N'$. Due to probability conservation,
the flow of probability is organized in closed cycles.

To illustrate things, we consider the marginal 
probability distribution
\begin{equation}
P(N_1,N_2,N_3) = \sum_{r_1=0}^{1}\sum_{r_2=0}^{1}\sum_{r_3=0}^{1}
	         P(N_1,N_2,N_3,r_1,r_2,r_3).
\end{equation}
Oscillatory behavior of the repressilator is characterized
by a regular cyclic pattern in the flow diagram 
$\phi(\vec N \to \vec N')$, as observed in
the marginal probability distribution.
In this diagram, the flow is from the $A$ dominated
region to the $C$ dominated region, then to the $B$ dominated
region and back to the $A$ dominated region.
Here we present results for a typical choice 
of parameters
for bacteria such as {\it E. coli}.
The values of these parameters are sensitive to the external
conditions, such as the temperature and the nutritional supply.
For a detailed list of parameters see Table 2.1
in Ref.
\cite{Alon2006} 
and Table 2
in Ref. 
\cite{Arkin1998}.
More specifically, 
the parameter values used here are
$g=0.05$, 
$d=0.003$, 
$\alpha_0=0.5$ 
and 
$\alpha_1=0.01$ 
(s$^{-1}$).
The protein synthesis rate $g$
represents typical synthesis times of proteins,
which are of the order of 10 to 20 seconds.
The degradation rate is consistent with
typical half-life times of proteins and mRNAs vary in the range
of several minutes
\cite{Alon2006,Arkin1998}.
The binding rate represents a time scale of diffusion across
the cell and specific binding of a transcription factor
to a DNA site, of the order of one second
\cite{Elowitz2000,Alon2006}.
The un-binding rate represents residence time on the 
DNA site of several minutes
\cite{Elowitz2000}.
It should be noted that the qualitative results are
not sensitive to the specific choice of the parameters.

\begin{figure}
\centering
\includegraphics[width=13cm]{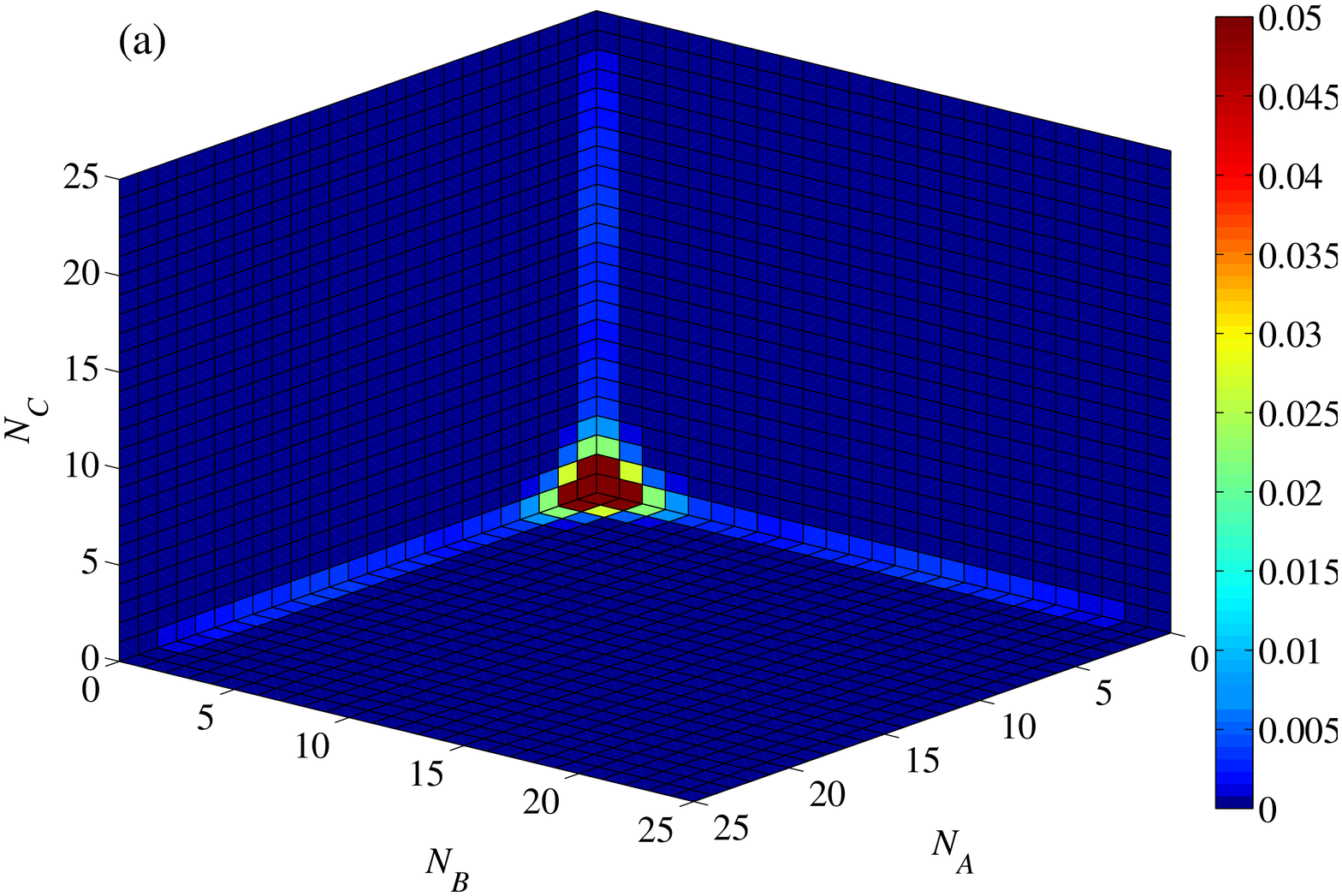}
\includegraphics[width=13cm]{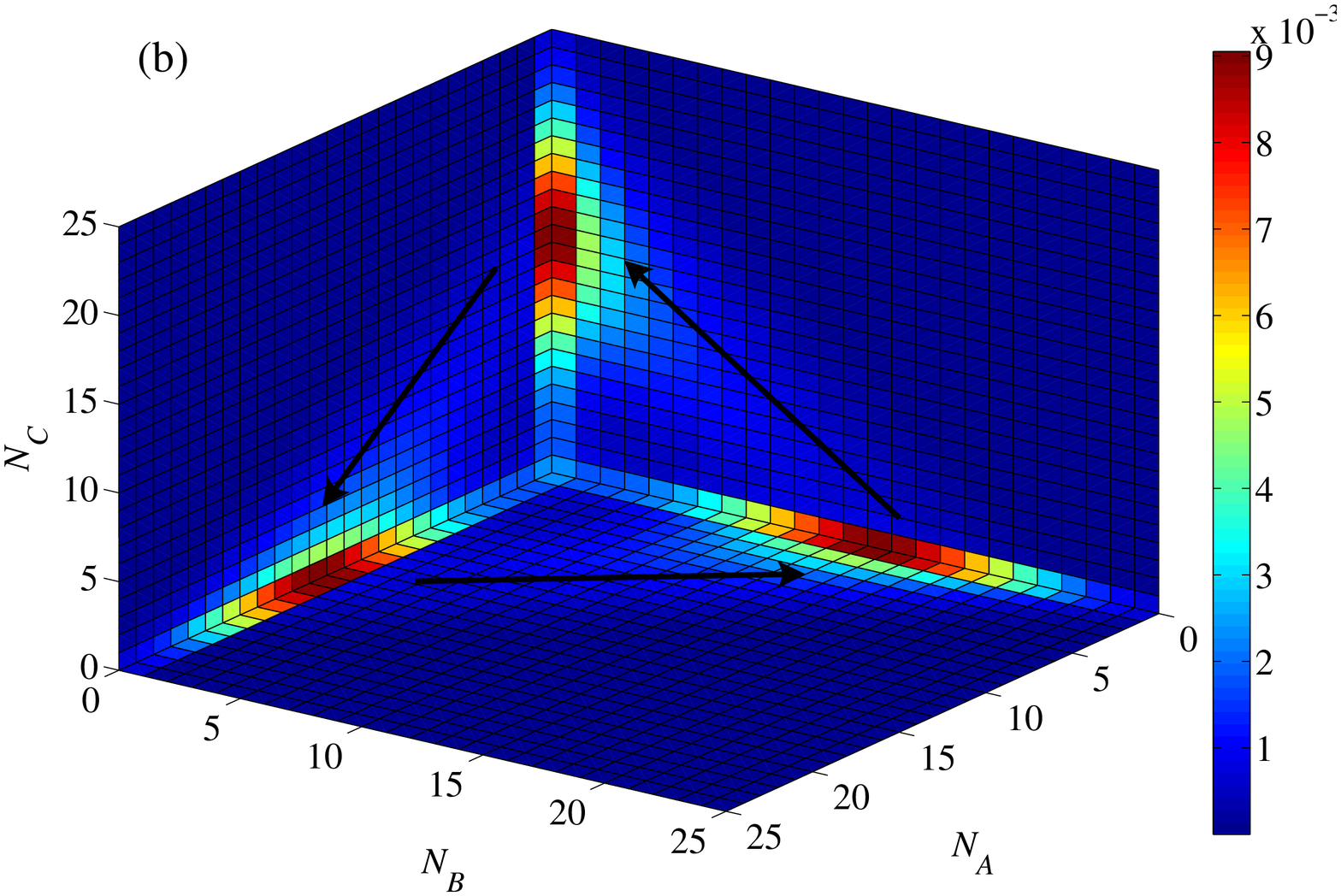}
\caption{
(color online) Slices of the probability distribution
$P(N_A,N_B,N_C)$ in the three planes $N_A=0$, $N_B=0$ and $N_C=0$
for the repressilator (the probability is represented by the color):
(a) without degradation of the bound repressors, 
where most of the probability is concentrated near the origin,
and there are no oscillations;
(b) with degradation of the bound repressors, where
$d_r=d=0.003$ (s$^{-1}$).
In this case the system exhibit oscillatory behavior.
Three peaks are observed, each of them dominated by one of 
the proteins. The probability flows between the peaks in the
directions marked by the arrows.
}
\label{fig:3}
\end{figure}

\begin{figure}
\includegraphics[width=13cm]{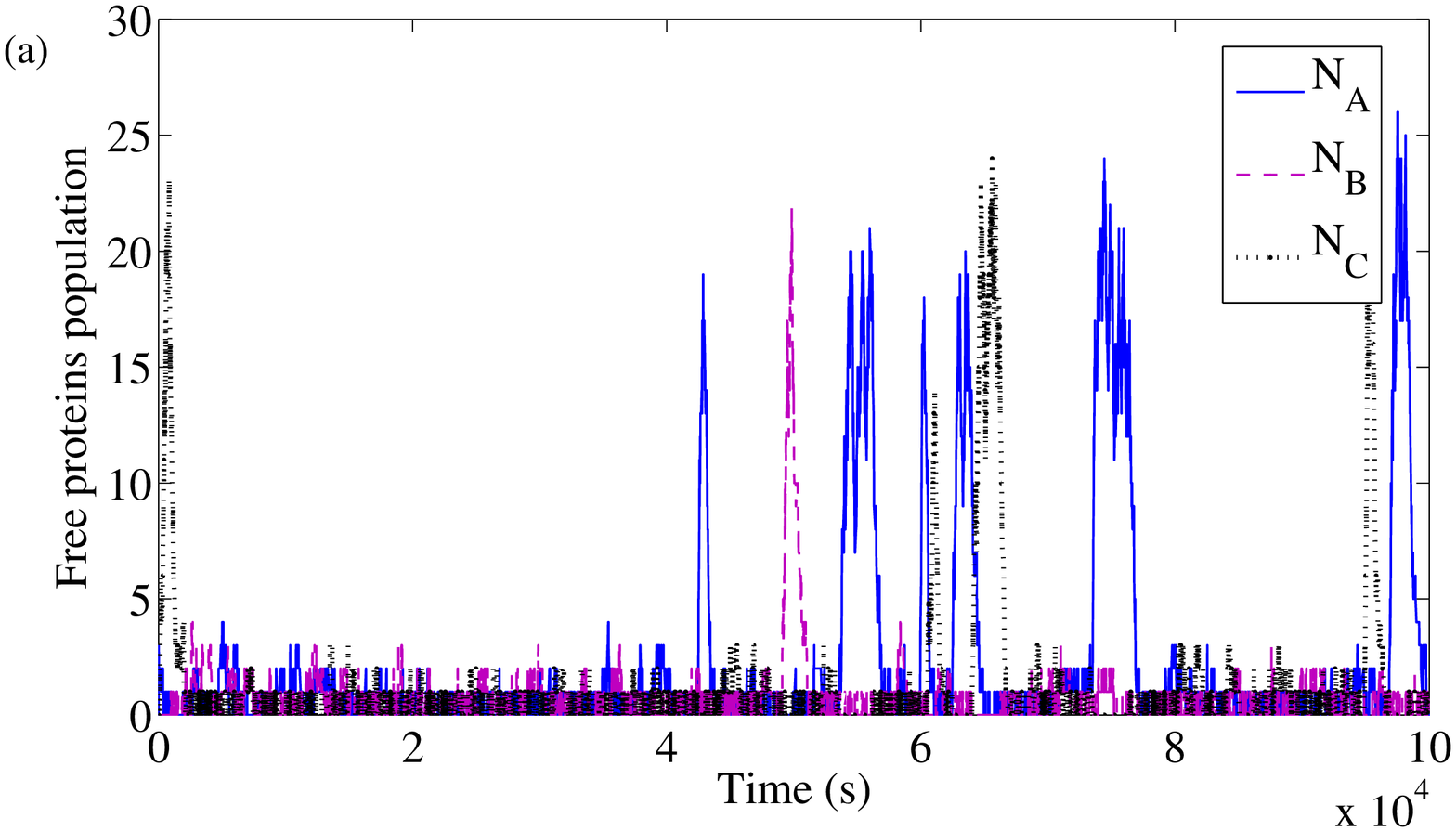}
\includegraphics[width=13cm]{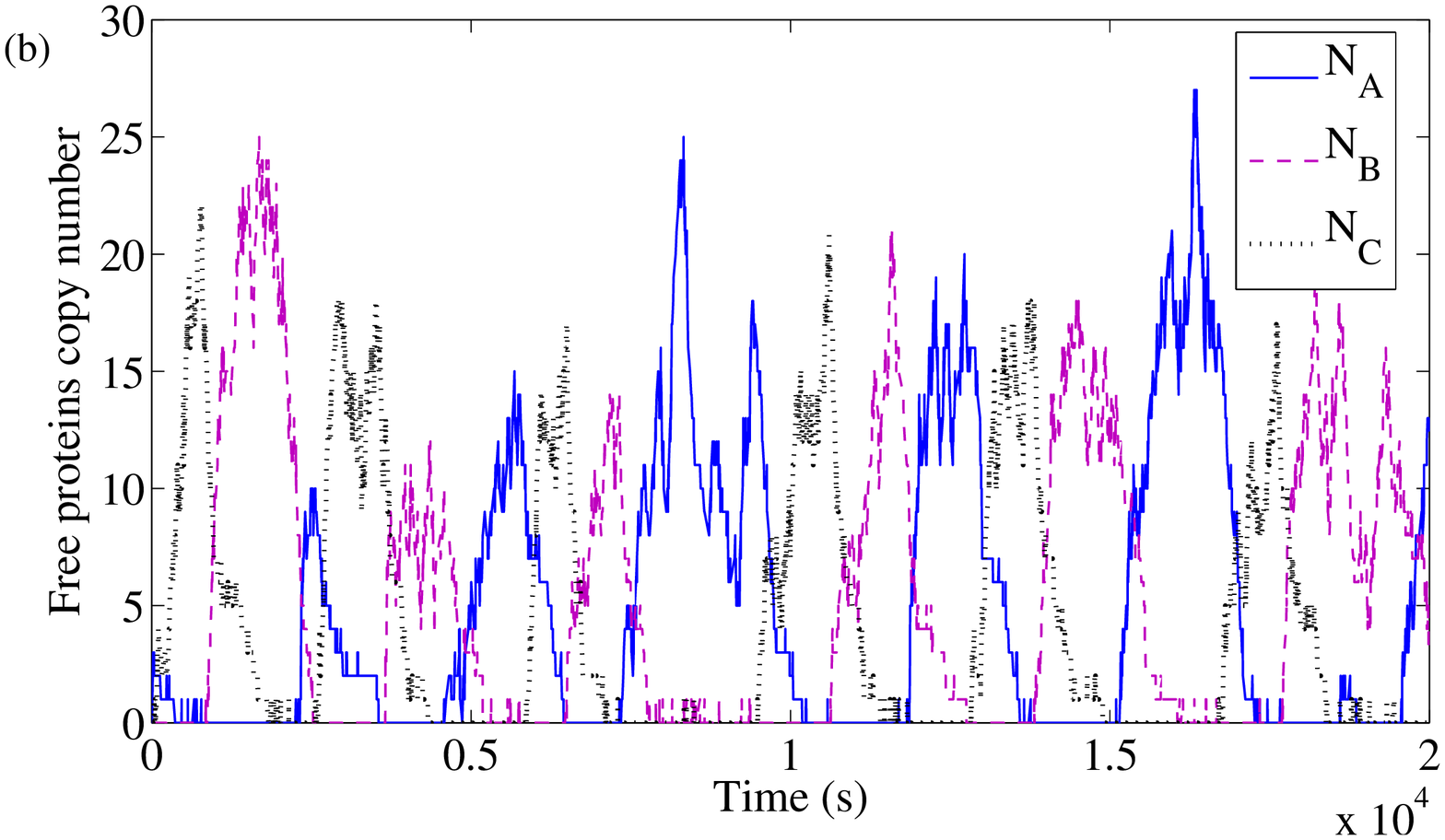}
\caption{
(color online)
The number of free proteins of types $A$, $B$ and $C$,
vs. time, for the repressilator,
obtained from MC simulations:
(a) without degradation of the bound repressors.
In this case all the proteins are suppressed most of the time,
with some irregular bursts;
(b) with degradation of the bound repressors,
where oscillations are observed.
The oscillations are irregular both in their
period and amplitude.
The average period is $\simeq 3750$ (s), which is significantly longer
than obtained from the rate-equations with the same parameters. 
The maximal number of proteins is also higher than in the 
rate-equation results.
}
\label{fig:4}
\end{figure}

In case that there is no degradation of bound repressors,
namely
$d_r=0$,
there are no oscillations.
The marginal probability distribution $P(N_A,N_B,N_C)$ 
is highly concentrated near the origin
[Fig.~\ref{fig:3}(a)]. 
In addition, there is a
small probability that one of the proteins 
will have a high copy number 
while the other two genes are not expressed.
We refer to this as a 'dead-lock' situation. 
The production of all the proteins is suppressed 
by the simultaneous binding of bound repressors,
and therefore oscillations cannot exist.
The probability flux is small and also concentrated near the origin.
MC simulations [Fig.~\ref{fig:4}(a)]
show that indeed, most of the time, all the proteins appear
in very low copy numbers 
(namely, two proteins or less).
Occasionally, there is a burst in the population of one of the proteins, 
but no regular oscillations are observed.

In order to obtain oscillations 
we introduce degradation of the bound repressors.
For simplicity, we assume that bound repressors degrade
at the same rate as free repressors, namely
$d_r=d=0.003$, leaving the other parameters as above.
This prevents the 'dead-lock' situation because degradation removes 
the bound repressors from the system. This is in contrast to un-binding, 
where the resulting free repressor may quickly bind again.
In this case oscillations are observed.
A similar effect was observed before in stochastic simulations
of the toggle switch
\cite{Lipshtat2006,Loinger2007}.
It was found that BRD induces bistability by removing
the dead-lock situation.

Under conditions in which oscillations take place,
the probability distribution, $P(N_A,N_B,N_C)$, exhibits three
different peaks [Fig.~\ref{fig:3}(b)].
Each peak represents the stage of the oscillation in which
the corresponding protein is dominant.
The peaks are connected through regions with smaller, but
non-vanishing probability. Through these regions probability flows
between the three peaks (see arrows).
MC simulations now show oscillatory behavior
[Fig.~\ref{fig:4}(b)].
In these oscillations $C$ domination is followed by $B$ domination,
then $A$ and returning to $C$.
However, the oscillations are not regular. 
Both the period and the amplitude
vary from one cycle to the next. 

\begin{figure}
\includegraphics[width=13cm]{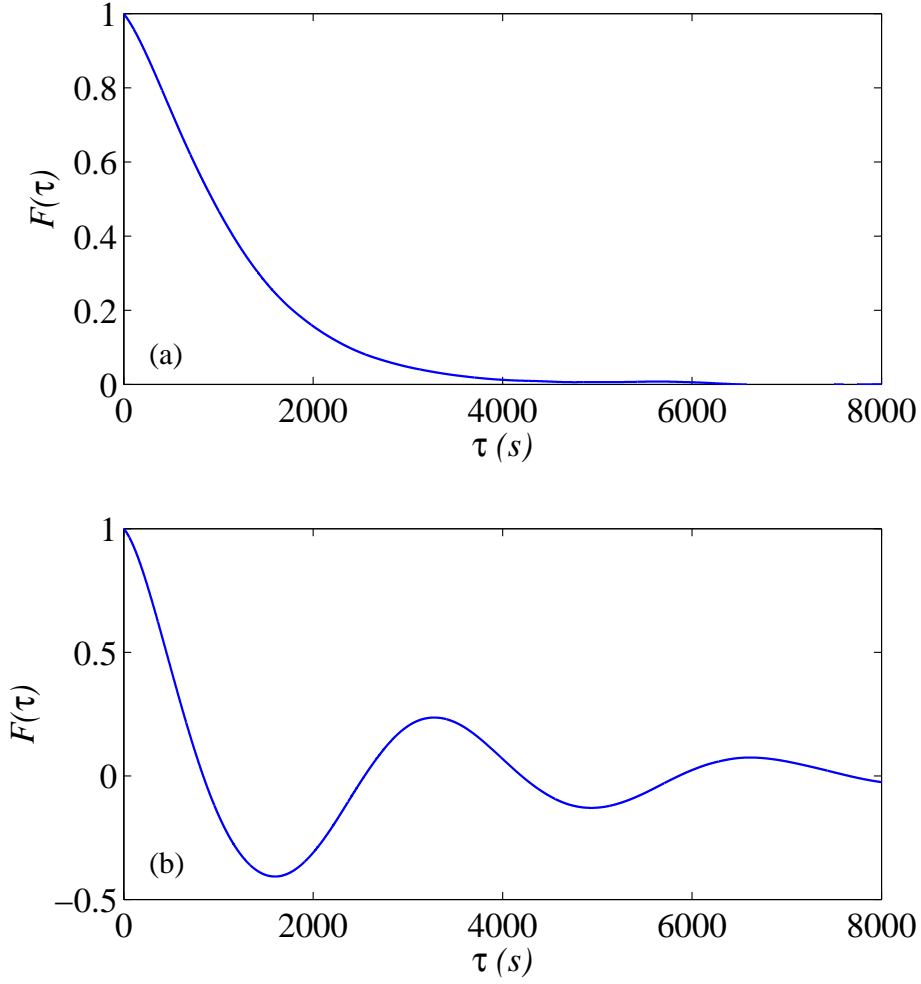}
\caption{
The Autocorrelation function $F(\tau)$ (normalized to unity),
computed for the output of MC simulations:
(a) in case there is no degradation of bound repressor ($d_r=0$).
In this case the system does not exhibit oscillations. 
This is indicated by the autocorrelation function decaying 
monotonically to zero;
(b) with degradation of bound repressor, $d_r=d=0.003$.
In this case the system exhibits (irregular) oscillations. 
This is indicated by the autocorrelation function which 
exhibits several oscillations before it decays.
}
\label{fig:5}
\end{figure}

Since in MC simulations the oscillations are not regular,
they are sometimes difficult to characterize.
In order to identify the oscillations we use the fact that oscillatory
systems exhibit a characteristic period, which can be evaluated 
using auto-correlation analysis.
The auto-correlation function is defined by

\begin{equation}
F(\tau)=\left<N_i(t+\tau)N_i(t)\right> - \left<N_i(t)\right>^2,
\end{equation}

\noindent
where $\left< \cdot \right>$ denotes averaging with respect to $t$.
When the system does not exhibit oscillations,
$F(\tau)$ decays monotonically to zero [Fig.~\ref{fig:5}(a)]. 
In case of oscillations, 
$F(\tau)$ oscillates before it decays
[Fig.~\ref{fig:5}(b)]. 
The location of the first maximum of $F(\tau)$ provides
the average period of the oscillations.
The phase coherence time is determined by the number
of the oscillations of $F(\tau)$ before it decays.

\subsection{Repressilator with Cooperative Binding}

In the deterministic analysis of this version of the circuit,
discussed in Sec. III B above, it was found that 
in order to obtain oscillations one needs to either
include the mRNA level or to assume bound-repressor degradation
(Table II).
MC simulations of the same circuit indicate that
in the stochastic case the situation is different. 
In this case the degradation of the bound repressors 
is a necessary condition for oscillations. 
The inclusion of the mRNA level does not affect the
appearance of oscillations in this case.

\begin{figure}
\includegraphics[width=13cm]{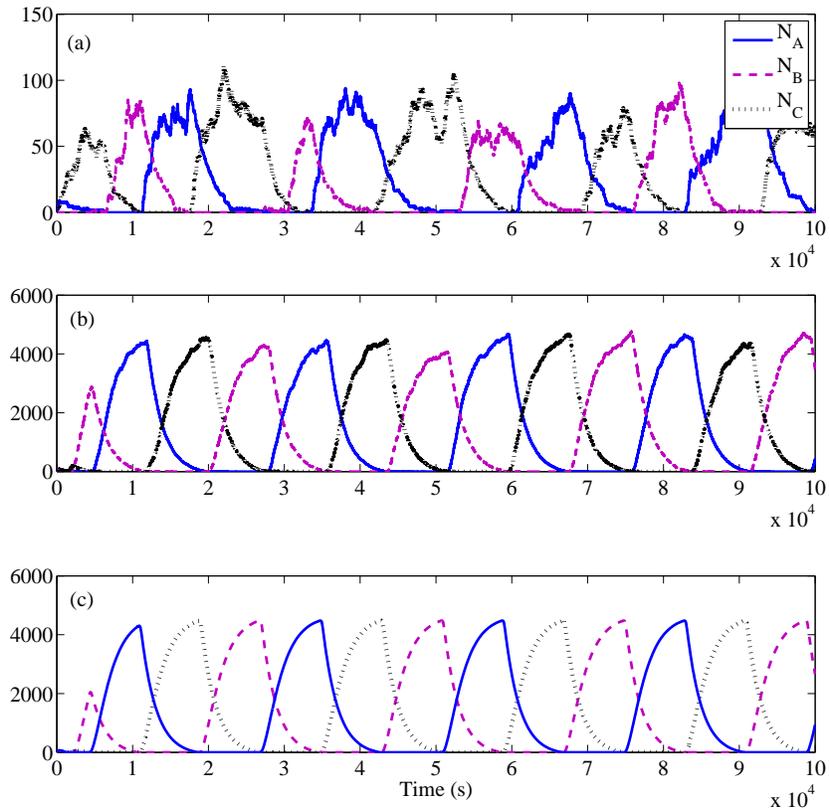}
\caption{
The number of free dimers, which consist 
of two proteins of types $A$, $B$ or $C$,
for the repressilator with cooperative binding.
(a) MC simulation results for a single plasmid;
(b) MC simulation results for 50 plasmids;
(c) Rate-equation results for 50 plasmids.
Note that as the number of plasmids increases the 
rate-equation and MC results become identical.
}
\label{fig:6}
\end{figure}

In 
Fig.~\ref{fig:6}(a)
we present the oscillations obtained from MC simulations 
of the repressilator with cooperative binding.
The mRNA level is included, in order to obtain a more
realistic description of the system.
The MC simulations are based on the master equation
for the probability 
$P(N_i,r_i,m_i,D_i)$, $i=1,2,3$,
for the cell to contain $N_i$ free proteins 
and $r_i$ bound proteins
of type $X_i$,
as well as 
$m_i$ copies of mRNA and 
$D_i$ copies of the corresponding dimer.
This master equation is not written explicitly here 
because it is cumbersome and adds little insight.
It can be reproduced by starting from
Eq. 
(\ref{eq:oscillator_BRD_master})
and adding the terms that correspond to the 
synthesis and degradation
of mRNAs as well as to dimer formation and degradation.
Due to the higher dimensionality of this equation, direct 
integration becomes infeasible and MC simulations are used.

\section{The effect of the number of binding sites}

We have examined the differences between the 
results obtained from deterministic and stochastic analysis
of the repressilator circuit.
We identified a case
in which oscillations are obtained only in the
rate-equations and are not obtained in MC simulations.
This is the case of 
the repressilator with cooperative binding and without BRD, 
where the mRNA level is taken into account explicitly
(Table II). 
Even when oscillations are obtained in both methods,
there are differences between them.
The oscillations obtained in the rate-equations 
are regular, and those
obtained from the MC simulations are noisy and irregular.
Moreover, the period and amplitude differ significantly 
between the rate-equations and the MC simulations.
Below we discuss and try to resolve these differences.

The rate-equations deal with continuous quantities.
These quantities are the averages, 
over an ensemble of cells,
of the actual copy numbers
of the proteins, which are discrete. 
The rate-equations involve some kind of 'mean field approximation'.
In general, this approximation is justified
when the copy numbers are large and the fluctuations can be ignored.
However, in our case, an essential part of the system, namely 
the bound repressors, always appear in small numbers, 0 or 1.
Therefore, the assumption of large copy numbers fails, and the
validity of the rate-equations is questionable.

The rate-equations can describe the system in a correct manner
only in the limit of high copy numbers of bound repressors.
Interestingly, this situation can, in fact, be realized in cells
by placing the relevant genes on plasmids, as done 
in Ref. 
\cite{Elowitz2000}.
Plasmids are small circular segments of DNA that may exist 
in the cell and can be inserted synthetically.
The number of plasmids in the cell, $n_p$, can be controlled.
The number of binding sites that regulate a particular gene,
which appears on the plasmids
is equal to $n_p$ 
if this gene does not appear on  the chromosome.
If it is also present on the chromosome, the number
of such binding sites is
$n_p+1$.
Here we assume, for simplicity,
that the number of the binding sites is $n_p$.
Taking this into account, 
appropriate changes must be made in the equations
describing the system.
The number of bound repressors,
$r_i$, can now take the values 
$0 \le [r_i] \le n_p$ in the rate-equations
and the values
$r_i = 0,1,\dots n_p$, 
in the master equation.
In both cases, the expression
$1 - r_i$
should be replaced by
$n_p - r_i$.
For example, 
Eq.~(\ref{eq:extended_rate}) 
becomes

%
\begin{equation}
\left\{
\begin{array}{l}
\dot{[X_i]} = g \cdot (n_p-[r_{i-1}])-d [X_i]
-\alpha_0[X_i]\left(n_p-[r_i]\right)
+\alpha_1[r_i] \\
\dot{[r_i]} = \alpha_0[x_i]\left(n_p-[r_i]\right)-\alpha_1[r_i].
\end{array}
\right.
\label{eq:extended_rate_with_plasmids}
\end{equation}

\noindent
In the limit
of a large number of plasmids, 
an agreement is obtained between the rate-equation
and the MC results.
This agreement is both qualitative and quantitative.
Qualitatively, for a high plasmid copy number,
the system exhibits oscillations in the
rate-equations if and only if it exhibits oscillations
in the MC simulations. 
Consider, for example, the repressilator with dimers
and without BRD,
where the mRNA level is taken into account.
For $n_p=1$ the system exhibits oscillations in the
rate-equations but not in the MC simulations.
As $n_p$ increases, the oscillations in the rate-equations 
disappear and become consistent with the
MC results. 

\begin{figure}
\includegraphics[width=12cm]{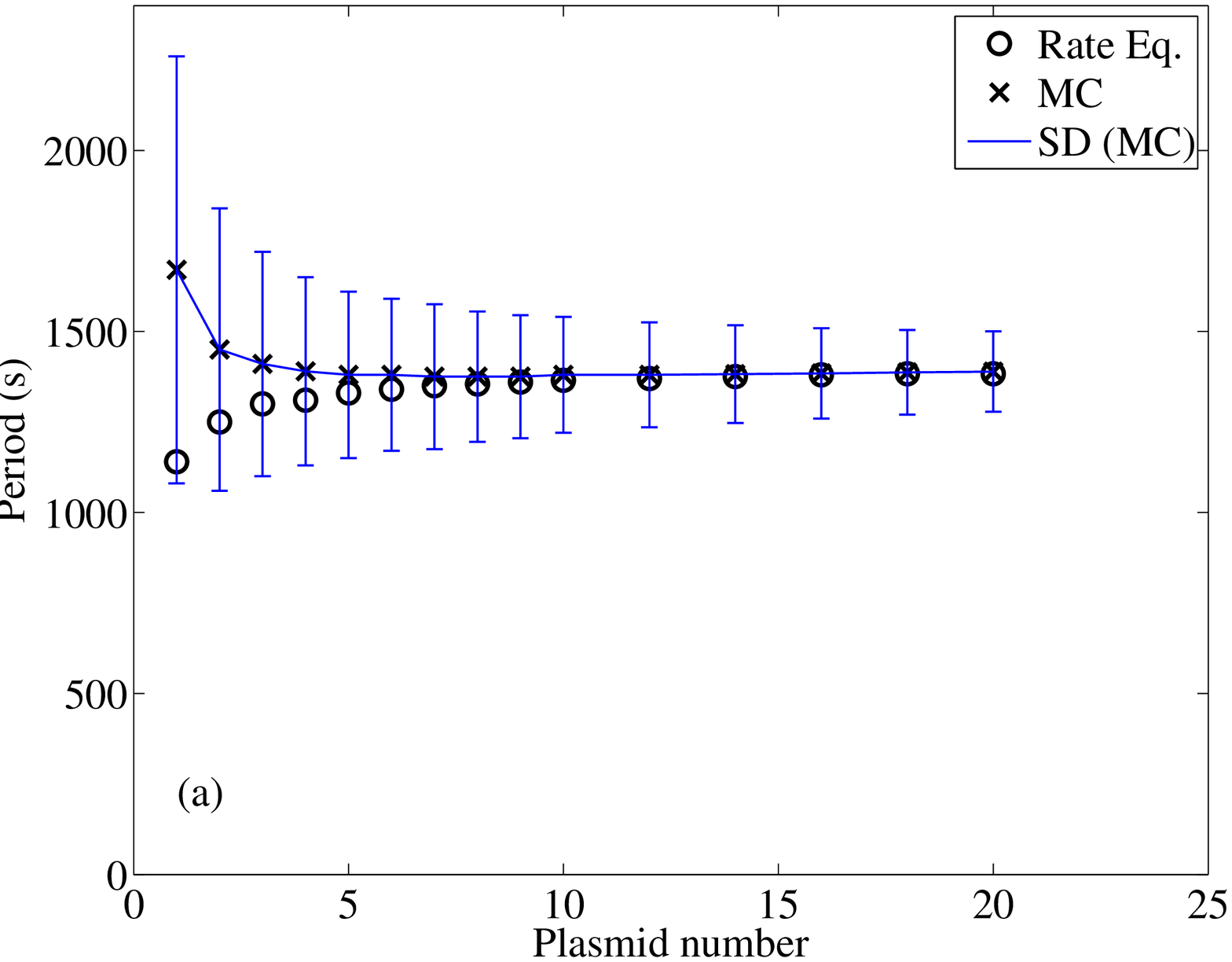}
\includegraphics[width=12cm]{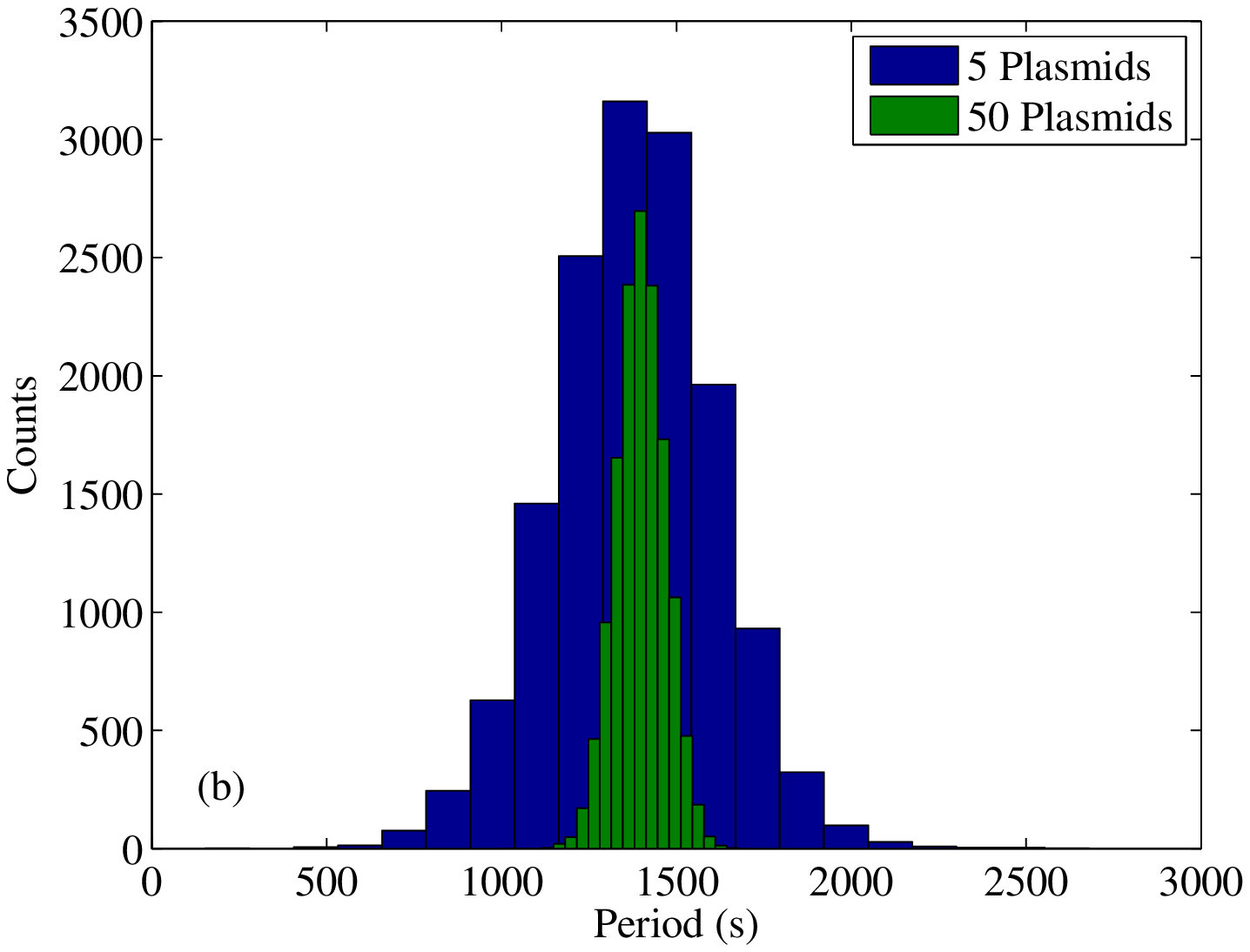}
\caption{
(a) Comparison between the period of the oscillations obtained
from MC simulations and from rate-equations for the same parameters,
vs. the plasmid copy number.
For a small number of plasmids the periods differ,
but for a large number of plasmids, they approach the same value.
The error bars represent one standard deviation (SD);
(b) The distribution of the periods of oscillations obtained from
MC simulations for 5 and 50 plasmids.
The distribution is approximately
a Gaussian with almost the same average.
The Gaussian is much narrower for the high plasmid copy number.
This indicates that the oscillations become more regular when
a large number of plasmids is present in the cell.
}
\label{fig:7}
\end{figure}

In case that the
number of plasmids is small, 
the average period 
of the oscillations in the 
MC simulations 
may differ from
the period obtained
in the rate-equations.
However, for a large number of plasmids, the oscillations
obtained in the MC simulations become much more regular, and more 
similar in shape to those obtained from the rate-equations, with
the same number of plasmids [Fig.~\ref{fig:6}].
In this case
the two periods converge towards each other
[Fig.~\ref{fig:7}(a)].
The distribution of the periods in the MC simulations 
becomes narrower as the number of plasmids increases
[Fig.~\ref{fig:7}(b)],
and the oscillations become more regular.

\section{Summary}

We have analyzed the
genetic repressilator 
circuit using deterministic and stochastic methods.
In particular, we examined the effects of 
cooperative binding, 
the degradation of bound repressors and
the inclusion of the mRNA level in the model.
The qualitative results are summarized in Table II.
Due to the small numbers of proteins and binding sites,
stochastic effects are significant and 
the deterministic analysis may fail.
It fails qualitatively in the biologically relevant
case in which there is cooperative binding,
the mRNA level is taken into account and
no BRD is assumed.
In this case the rate-equations predict oscillations
which do not appear in the stochastic analysis.
In addition, even when the deterministic 
and stochastic methods agree about the existence
of oscillations, there are quantitative differences
in the period, amplitude and regularity of the
oscillations as well as in the range of parameters
in which they appear.

Since the repressilator was encoded on plasmids we have
studied the effect of increasing
the number of plasmids in a cell
on the behavior of the system.
We found that as the number of plasmids increases,
the role of fluctuations is suppressed and the rate-equations
become valid.
The results show that varying the plasmid copy number may lead to 
qualitative changes in the dynamics of genetic circuits.
This prediction can be tested experimentally in the
context of synthetic biology 
and should be taken into account in the
design of artifical genetic circuits.

Our results indicate that deterministic analysis is valid only
in the limit in which all the components, namely mRNAs and both
free and bound proteins, appear in large copy numbers.
This condition is not satisfied in genetic circuits
encoded on the chromosome.
Thus, for these circuits deterministic methods may fail.
In particular, in cases in which the system exhibits 
multiple steady states
\cite{Lipshtat2006,Loinger2007}
or oscillations, deterministic
and stochastic methods may yield qualitatively different
results.
In these cases, the system may be sensitive to subtle
features such as cooperative binding, BRD and the inclusion 
of the mRNA level in the model.
Thus, in the modeling of these systems, such features
should be taken into account.
In addition, our results provide strong evidence for the
existence of degradation of bound proteins.
This result has significant biological implications
beyond the specific circuit studied here.

We thank N.Q. Balaban for many helpful discussions.

\newpage
\clearpage

\begin{table}
\caption{
The existence of oscillations vs. the Hill-coefficient $n$
in the Michaelis-Menten equations, with and without
the inclusion of the mRNA level.
The mRNA level is included in Eq. (\ref{eq:michaelis_menten_with_mrna})
and is not included in Eq. (\ref{eq:michaelis_menten_without_mrna}).
In both cases, we did not include bound-repressor degradation.
The cases in which the system exhibits oscillations are marked by $\surd$
and those in which it does not are marked by $\times$.
}

\begin{tabular}{ccc}
\hline \hline
Hill-coefficient \, &\, with mRNA \,&\, without mRNA \\ \hline
1 & $\times$ & $\times$ \\
2 & $\surd$  & $\times$ \\ 
3 & $\surd$  & $\surd$ \\ \hline
\end{tabular}
\end{table}

\begin{table}
\caption{
Oscillations in different variants of the repressilator.
The cases in which the system exhibits oscillations are marked by $\surd$
and those in which it does not are marked by $\times$.
The following features are taken into account: 
cooperative vs. non-cooperative regulation, the inclusion vs.
non-inclusion of the mRNA level in the model and 
degradation vs. non-degradation of bound repressors.
Here, cooperative circuits refer to 
repression by protein dimers.
The deterministic analysis is done using the complete set 
of rate-equations and the stochastic analysis is done using
MC simulations.
Note that in cases in which both the deterministic and stochastic
approaches exhibit oscillations, the parameter range 
in which they appear may differ.
In the limit of high plasmid copy number, the results obtained
from the deterministic and stochastic method coincide.
The results reported in this Table are based on
linear stability analysis and on a large number of 
simulations covering the biologically relevant range of the
parameter space with a fine grid.
}

\begin{tabular}{|c|cc|cc|c|}
\hline \hline
\multicolumn{3}{|c}{Circuit variant} \, & 
\multicolumn{2}{|c|}{Low plasmid copy number} &\, 
High plasmid copy number \\  \hline
  & mRNA & BRD \, &\, Deterministic \,&\, Stochastic  \,& Deterministic and Stochastic\, \\ \hline
 & No & No & $\times$ & $\times$ & $\times$      \\
 & Yes & No &  $\times$ & $\times$ & $\times$      \\
\raisebox{2.5ex}[0pt]{Non-Cooperative} 
& No & Yes &  $\surd$ & $\surd$ & $\surd$             \\
 & Yes & Yes & $\surd$ & $\surd$ & $\surd$      \\ \hline
 & No & No & $\times$ & $\times$ & $\times$      \\
 & Yes & No & $\surd$ & $\times$ & $\times$      \\
\raisebox{2.5ex}[0pt]{Cooperative} 
& No & Yes &    $\surd$ & $\surd$ & $\surd$     \\
 & Yes & Yes & $\surd$ & $\surd$ & $\surd$      \\ \hline
\end{tabular}
\end{table}

\newpage
\clearpage

\end{document}